# Direct observation of magnetocaloric effect by differential thermal analysis: influence of experimental parameters

Yamila Rotstein Habarnau[a,b], Pablo Bergamasco[a,b], Joaquin Sacanell[a], Gabriela Leyva[a,c], Cecilia Albornoz[a], Mariano Quintero[a,c]

a Departamento Materia Condensada, CAC, CNEA, Av. Gral Paz 1429, San Martin (1650), Argentina

b Departamento de Física, FCEN, Universidad de Buenos Aires, Ciudad Universitaria Pab.I, Buenos Aires (1428), Argentina,

c Escuela de Ciencia y Tecnología, UNSAM, San Martín, Buenos Aires (1650), Argentina

*Abstract*

The magnetocaloric effect is the isothermal change of magnetic entropy and the adiabatic temperature change induced in a magnetic material when an external magnetic field is applied. In this work, we present an experimental setup to study this effect in metamagnetic transitions, using the differential thermal analysis technique, which consists in measuring simultaneously the temperatures of the sample of interest and a reference one while an external magnetic field ramp is applied. We have tested our system to measure the magnetocaloric effect in $La_{0.305}Pr_{0.32}Ca_{0.375}MnO_3$, which presents phase separation effects at low temperatures (T < 200 K). We obtain ΔT vs H curves, and analyze how the effect varies by changing the rate of the magnetic field ramp. Our results show that the intensity of the effect increases with the magnetic field change rate.

We also have obtained the effective heat capacity of the system without the sample by performing calorimetric measurements using a pulse heat method, fitting the temperature change with a two tau description. With this analysis, we are able to describe the influence of the environment and subtract it to calculate the adiabatic temperature change of the sample.

**Introduction.**

Since de observation of magnetocaloric effect (MCE) near room temperature in 1997 by Pecharsky et al [i] an important increase in the amount of work in that topic has been observed. Most of the work has been devoted to obtain a large MCE around room temperature using the most stable, cheapest and easy-to-obtain material.

In that sense, materials belonging to different kinds, such as intermetalic compounds, rare earth metals and alloys and manganites were proposed as possible alternatives. But, as the complexity of the systems growths, a more careful interpretation of the relation between the magnetization results and the MCE is needed.

The most accepted approach to study MCE is the use of a Maxwell relation to connect the magnetization of the sample with the entropy change induced by the magnetic field.

An alternative approach has been proposed by Giguère[ii] *et al.* and Balli *et al*[iii] using a Clausius-Claperyon equation to obtain a more realistic calculation of the magnitude of the MCE during a metamagnetic phase transition. The thermodynamic equivalence of both approaches has been demonstrated by Sun et. al.[iv] Despite of that, the results obtained not always are the same.

One example of this is the controversy over study of the MCE in the "Mn1-xFexAs" system, where an incorrect use of a thermodynamic model results in an overestimation of the entropy change during the metamagnetic transition[iii].

Another alternative to the description of the MCE considers the enthalpy change of the system during the metamagnetic transition and has been supported with direct measurements even for inverse MCE[v].

The above presented scenario reveals the necessity of an effort from the scientific community pointing to understand how the MCE reveals in the most usually measurable quantities, such as magnetization and resistivity. One of the first limitations of this kind of work is the absence of an easy way to perform a direct measurement of the MCE. As consequence, from the large number of publications studying MCE, only a small fraction contains direct measurements of the effect.

In this work we present a detailed study of the MCE using the differential thermal analysis (DTA) technique. The influence of the experimental parameters, such as temperature, external pressure and rate of change of magnetic field will be presented and discussed. Finally, the results of DTA will be described using a thermodynamic model from magnetization measurements.

**Experimental details**

We have developed an experimental setup to perform DTA measurements using a commercial system (Versalab–Quantum Design$^{TM}$). The system consist of two Pt resistances mounted on the Versalab sample holder, adapted for DTA measurements, as schematically shown in **Fig. 1**. The compound under study and a reference sample (a piece of alumina) are in thermal contact with each of the Pt resistances. A teflon layer separates the Versalab puck and the thermometers to optimize thermal insulation of the sample and the reference. We perform a simultaneous measurement of both thermometers to obtain the local temperature of each sample as the magnetic field is applied. The differential analysis is performed to get rid of any magnetoresistance within the Pt sensor and small thermal fluctuations, common to both thermometers.

The sample used to test the system is a ceramic $La_{0.305}Pr_{0.32}Ca_{0.375}MnO_3$ manganite (LPCMO), which presents metamagnetic transition between antiferromagnetic and ferromagnetic phases at temperatures below 200 K[vi]. A typical curve of DTA is presented in figure 2, with the corresponding magnetization loop at 102 K.

When the magnetic field is increased from 0 to 3 T the sample´s temperature increases with a peak in DT at H=17000 Oe. At the same field the system undergoes from a CO to FM state through a metamagnetic transition. While decreasing H, the opposite transition is observed and the sample temperature decreases with a negative $\Delta T$ peak around 3000 Oe.

The shape of the DTA curve is a direct consequence of the MCE presented in the sample, but it is also influenced by many experimental parameters. In what follows, we present a careful analysis of the effect of these parameters. The thermodynamics of the MCE analyzes the effect considering the adiabatic temperature change related with the application of the external magnetic field. In the case of a direct measurement of the effect, is clear that the transition is far away from being adiabatic. The sample exchange its heat with the thermal bath and the different instruments for measurement. As a consequence the measured ΔT is quite below the adiabatic ΔT.

Proposing the heat balance equations of the system schematically represented in figure 1 under the following considerations:
1) the thermal coupling with the thermal bath is almost the same for the sample, reference and sensors ($\gamma_E$)
2) the sample and the reference are in good thermal coupling with the corresponding sensors ($\gamma_{RS}, \gamma_{SS} >> \gamma_E$)

it is possible to obtain an equation for the temperature difference between the sample and the reference.

Defining $\Delta T = T - T_R$ where T and $T_R$ are the sample and the reference temperatures, respectively and supposing that the adiabatic temperature change is

$$dT_{AD} = \xi(H)dH \quad \text{(eq1)}$$

then $\Delta T$ is a solution of the following equation

$$(C_{Pt} + C_S)\frac{d\Delta T}{dt} = -2\gamma_E \Delta T - \xi(H)\frac{dH}{dt} + \frac{dT_R}{dt}\Delta C \quad \text{(eq 2)}$$

where $C_{Pt}$ and $C_S$ are the heat capacity of the sensor and the sample respectively. In this equation, the heat sources for the $\Delta T$ change are the coupling with the thermal bath and the MCE (first and second terms in the right side respectively). The last term represents the temperature difference associated with the asymmetries between the sample and the reference, being $\Delta C = C_{Ref} - (C_{Pt} + C_S)$. By an appropriate choice of the reference, this contribution can be reduced, allowing to neglect the last term.

From numerical integration of the equation 2 using the measured DTA curve is possible to obtain the adiabatic temperature change ($\xi(H)$) that contain the thermodynamic information of the system under study.

**Results**
As has been showed in equation 2, the DTA curve strongly depends of the experimental parameters. In the following we will show how is this dependence, and how the heat capacity of the sensors and reference can be measured to allow the determination of $\xi(H)$.

*Magnetic field rate*
When the rate of change of the magnetic field is modified, important variations in the intensity of the magnitude of the DTA curve are observed. In figure 3 we show an example of this variation, with field rates between 50 and 300 Oe/sec.
The effect is related with the heat exchange between the sample and the thermal bath (characterized by $\gamma_E$ (figure 1b). An increase in the magnetic field rate inhibits the temperature relaxation of the sample to the thermal bath temperature. In the inset of figure 3 we show the dependence of $\Delta T_{MAX}$ as function of the magnetic field rate. An almost linear behaviour is observed and the magnitude of the effect can be duplicated increasing the rate from 50 to 200 Oe/sec.

*System Heat Capacity*
To take into account the heat capacity of the sensors and the reference, a previous measurement is performed without the sample.

The method used in this case was the heat pulse where a high current pulse is applied on the Pt-1000. Measuring the temperature response of the system during and after the pulse [vii] is possible to extract the contribution of the system to the total heat capacity.

In figure 4 we show the effective heat capacity of both sensors. The obtained values (between 20 and 80 mJ/K) are comparable with the sample heat capacity (30-150 mJ/K), being this indicative of the importance of an adequate modification of the samples $\Delta T$ to obtain the adiabatic $\Delta T$.

**Conclusions**
In this work we presented a detailed characterization of a DTA system mounted on a commercial measurement system to analyze the MCE effect during metamagnetic transitions. We focused our attention in the environmental effects associated with the use of the technique and the way to get rid of them.
The dependence of the results with the rate of change of the magnetic field has been studied. We observed an important increase in the DTA signal as the rate of change of the magnetic field increases, reducing the heat exchange between the sample and the thermal bath.

Finally, to obtain a correct value for the adiabatic temperature change, the effective heat capacity has been established using a standard heat pulse technique.

We thank F. Parisi for helpful discussion of the results. J. Sacanell and M. Quintero are also members of CIC-CONICET.

**Figure captions:**

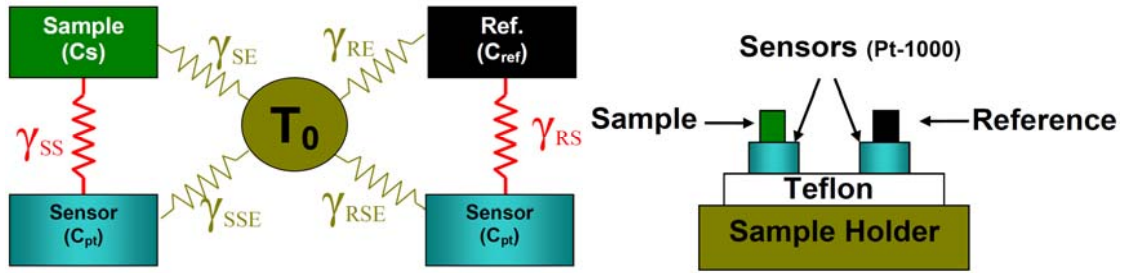

Figure 1: a) Representation of the DTA system with the thermal links associated with the different sources of heat exchange. b) Scheme of the experimental setup.

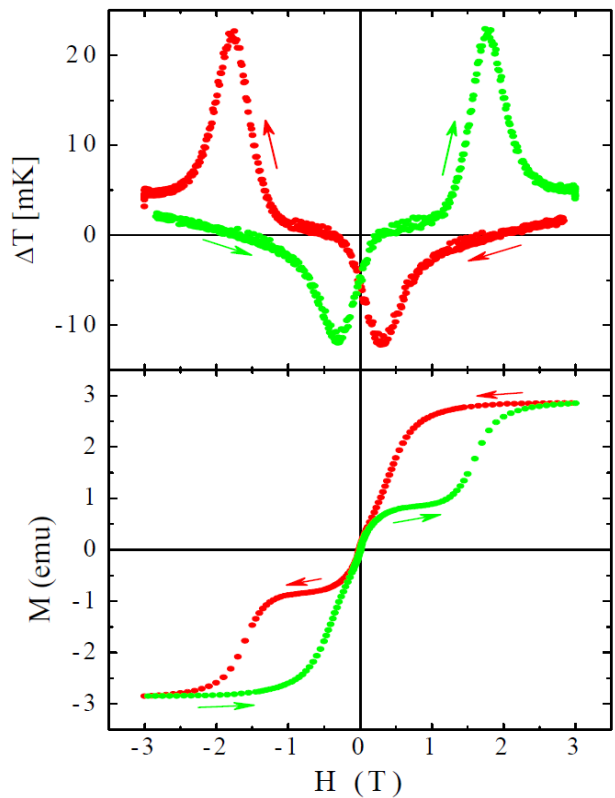

Figure 2: $\Delta T$ and Magnetization as function of magnetic field for T = 102 K in the sample LPCMO 0.32. The arrows indicate the sense of the applied field.

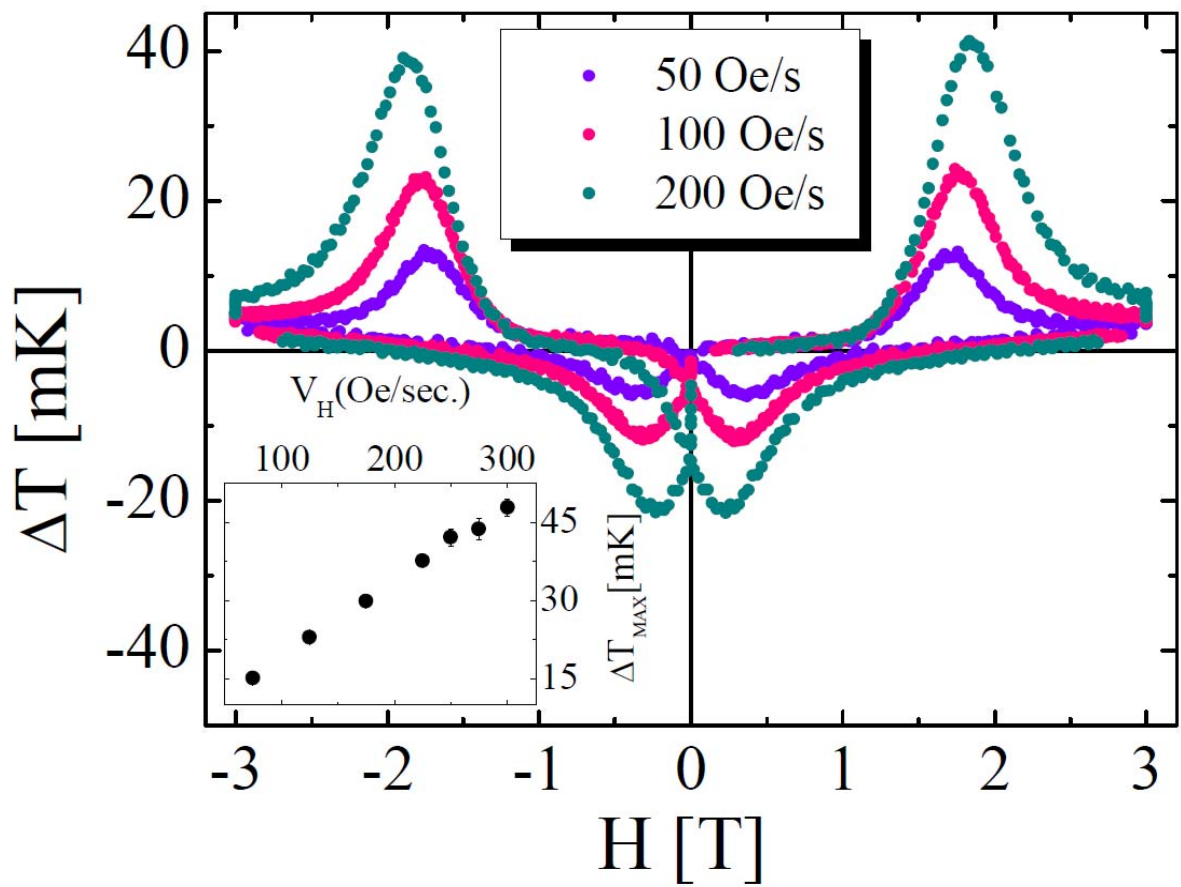

Figure 3: ∆*T* curve with different rates of applied magnetic field. Inset: DTmax vs Velocity

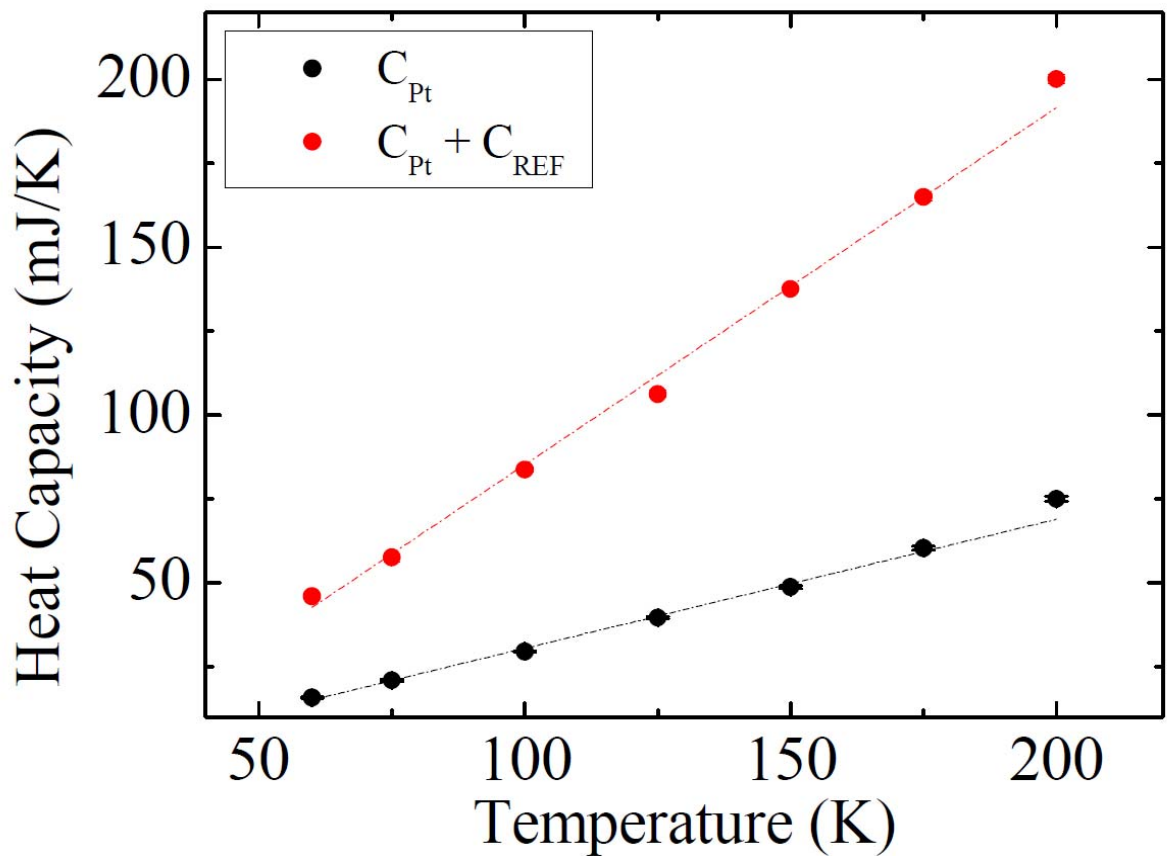

Figure 4: heat capacity of the sample sensor and the reference sensor.

---

[i] V. K. Pecharsky and K. A. Gschnerdner, Jr., Phys. Rev. Lett. 78, 4494 _1997_.

[ii] A. Giguère, M. Foldeaki, B. Ravi Gopal, R. Chahine, T. K. Bose, A. Frydman, and J. A. Barclay, Phys. Rev. Lett. 83, 2262 _1999_.

[iii] M. Balli, D. Fruchart, D. Gignoux, and R. Zach, Appl. Phys. Lett. 95, 072509 _2009_.

[iv] J. R. Sun, F. X. Hu, and B. G. Shen, Phys. Rev. Lett. 85, 4191 _2000

[v] M. Quintero, J. Sacanell, L. Ghivelder, A. M. Gomes, A. G. Leyva, and F. Parisi, Appl. Phys. Lett. 97, 121916 (2010)

[vi] M. Quintero, G. Leyva, P. Levy, F. Parisi. O. Agüero, I. Torriani, M.G. das Virgens, L. Ghivelder, Physica B, Vol. 354, Issues 1-4, P. 63-66 (2004).

[vii] Jih Shang Hwang, Kai Jan Lin, and Cheng Tien, **Rev. Sci. Instrum. 68 (1), January 1997**